\documentclass[prx, twocolumn, showkeys]
{revtex4-2}
\usepackage{graphicx} 
\usepackage{amsmath}
\usepackage{physics}
\usepackage{amsfonts}
\usepackage{amsthm}
\usepackage[ruled,lined]{algorithm2e}
\usepackage{multirow} 
\usepackage{algpseudocode}
\usepackage{bm}
\usepackage{times}
\usepackage{amsbsy}
\usepackage{enumitem}
\usepackage{color}
\usepackage{dcolumn}
\usepackage[utf8]{inputenc}

\usepackage{lipsum}
\usepackage{subfigure}
\usepackage{caption}

\usepackage[english]{babel}
\begin{document}
\theoremstyle{plain}
\newtheorem{theorem}{Theorem}
\newtheorem{lemma}[theorem]{Lemma}
\newtheorem{corollary}[theorem]{Corollary}
\newtheorem{proposition}[theorem]{$\eta_{r}$oposition}
\newtheorem{conjecture}[theorem]{Conjecture}

\theoremstyle{definition}
\newtheorem{definition}[theorem]{\gammaefinition}

\theoremstyle{remark}
\newtheorem*{remark}{Remark}
\newtheorem{example}{Example}

\title{Quantum-inspired attribute selection algorithm: A Fidelity-based Quantum Decision Tree}

\author{Diksha Sharma$^{1}$}
\author{Parvinder Singh$^{2}$}
\email{parvinder.singh@cup.edu.in}
\author{Atul Kumar$^{1}$}
\email{atulk@iitj.ac.in} 
\affiliation{$^{1}$Indian Institute of Technology Jodhpur, Rajasthan-342037, India}
\affiliation{$^{2}$Central University of Punjab Bathinda, Punjab-151001, India} 
\date{\today}
\begin{abstract}
A classical decision tree is completely based on splitting measures, which utilize the occurrence of random events in correspondence to its class labels in order to optimally segregate datasets. However, the splitting measures are based on greedy strategy, which leads to construction of an imbalanced tree and hence decreases the prediction accuracy of the classical decision tree algorithm. An intriguing approach is to utilize the foundational aspects of quantum computing for enhancing decision tree algorithm. Therefore, in this work, we propose to use fidelity as a quantum splitting criterion to construct an efficient and balanced quantum decision tree. For this, we construct a quantum state using the occurrence of random events in a feature and its corresponding class. The quantum state is further utilized to compute fidelity for determining the splitting attribute among all features. Using numerical analysis, our results clearly demonstrate that the proposed algorithm cooperatively ensures the construction of a balanced tree. We further compared the efficiency of our proposed quantum splitting criterion to different classical splitting criteria on balanced and imbalanced datasets. Our simulation results show that the proposed splitting criterion exceeds all classical splitting criteria for all possible evaluation metrics.
\end{abstract}
\maketitle
\section{Introduction}
The decision tree algorithm emulates human behavior to make a conclusion out of the provided information. It constructs an upside-down tree-like structure by placing the crucial attribute at the root node and recursively reaches at leaf node to draw a conclusive remark out of the provided dataset. The pivotal point for constructing a decision tree is its splitting criterion, which optimally segregates the whole dataset. Conventionally, there are two main splitting criteria- information gain and gini index. Both of these splitting criteria select a feature using their respective measures to divide the whole dataset into sub-datasets for developing a tree-like structure. Information gain uses Shannon's entropy to extract the stored information between a feature and the class attributes and recursively divides the dataset on a feature corresponding to the highest gain value \cite{quinlan1986induction}. Whereas, gini index uses conditional probability of a misclassified instance and therefore chooses the lowest value to split a dataset \cite{breiman2017classification}. Recently, the applications of quantum computing are showing its importance in diverse domains including machine learning. The fundamental laws of quantum computation, entanglement, and quantum parallelism are observed to be evolving sub-routines of machine learning algorithms. The fusion of classical machine learning and quantum computing has shown significant benefits in supervised and unsupervised algorithms such as quantum support vector machines \cite{rebentrost2014quantum}, quantum k-nearest algorithms \cite{schuld2014quantum}, k-means clustering algorithms \cite{sarma2019quantum}, and also for quantum neural networks \cite{kak1995quantum}. In the nutshell, quantum computation assists machine learning in the refinement of space and time complexities. \par
In the context of quantum decision trees, Farhi and Gutman \cite{farhi1998quantum} proposed  formulating computational problems into a decision tree that can be further penetrated using time-independent Hamiltonian by evolving a quantum state through nodes of decision tree to reach \textit{n}th node or solution. They further claimed that decision trees can be penetrated exponentially faster by quantum evolution than classical random walk. On similar lines, Burhman and Wolf \cite{buhrman2002complexity} proposed that a complex problem can be subdivided into multi-sub-problems, where the \textit{i}th query depends on the previous queries. This process can be thought of as a decision tree where the complexity of a problem is determined by depth of the tree. They computed a boolean function by making queries such that the computational complexity of the function is determined by a minimum number of queries required to make on any input. They also computed the complexity using deterministic, randomized and quantum computation. However, the authors find no benefit of one approach over the other. Therefore, these two studies used quantum computation to solve complex problems which can be sub-divided into the form of a decision tree. Later, Lu and Braustein proposed the use of von Neumann entropy as a splitting criterion for quantum decision tree \cite{lu2014quantum}. For this, they constructed a quantum state where amplitude values correspond to feature values and similarly generated a quantum state for class attribute, however, the attribute selection criterion requires a much more involved analysis \cite{schuld2015introduction}. Heese \textit{et al.} \cite{heese2022representation} proposed the idea of using a classical binary decision tree to construct a quantum circuit utilizing a genetic algorithm. The constructed quantum circuit, also called a Q-tree, is used for determining the class label for query data by measurement of the quantum circuit. However, it requires qubits as per the depth+features+labels of a classical decision tree and numerous measurements. Khadiv \textit{et al.} \cite{khadiev2019quantum} proposed a quantum version of the classical C5.0 decision tree algorithm. As per the proposal, searching for a maximum or minimum value corresponding to features is the only factor that can further enhance the time complexity of the classical decision tree and therefore they replaced the searching algorithm with Durr and Hoyer's minimum search algorithm \cite{durr1996quantum} to refine the time complexity. Further, Khadiv and Safina \cite{khadiev2022quantum} suggested using an amplitude amplification algorithm based on, Grover's search algorithm \cite{grover1996fast} to predict the class label for an unknown data point out of an ensemble of machine learning classifiers. Although these algorithms proposed different approaches, there is no formulation for the quantum decision tree that considers the evolution of balanced tree starting from a classical dataset. In this work, we propose a quantum splitting criterion based on fidelity and compare its efficacy with classical information gain and gini index criteria.
\section{Motivation}
As compared to other classical supervised algorithms, a decision tree is used widely for classification problems because of its ability to work with different measurement scales and therefore classifies a dataset including discrete and continuous features \cite{pal2003assessment}. It efficiently predicts the correlation between a feature and class without any prior statistical calculations. Moreover, the decision tree algorithm is interpretive and greedy in nature. Considering these advantages, it has applications in anomaly detection \cite{singh2018decision}, feature engineering \cite{piramuthu1994using}, priority prediction and is used as a basic building block for ensembling techniques \cite{dietterich2000experimental}. However, the greedy nature of the decision tree algorithm also leads to selecting a feature that ends up creating an imbalanced tree. An intuitive approach is to select an attribute that has a maximum correlation between occurrence of random events of a feature and corresponding class labels. \par
In this scenario, the fundamentals of quantum computing such as examining the correlation among states and the probabilistic nature can assist in achieving a balanced and efficient algorithm for the decision tree. Although there are a few instances of algorithms in  context of quantum decision trees, the authors either proposed the idea to approach a complex problem by formulating it as a classical decision tree and traversing it using quantum computing or refined the time complexity of classical decision algorithms using quantum algorithms. Here, we propose an efficient algorithm to construct and analyze a quantum decision tree for balanced as well as imbalanced datasets.For this, we propose to use fidelity as a measure for choosing a feature that optimally splits a dataset. In order to analyze our algorithm, we first define a two-qubit quantum state from the classical dataset using correlations between occurrence of class labels and the corresponding random events of a feature. The constructed quantum state is further utilized to compute fidelity between a feature and class attributes. This procedure of choosing a splitting feature includes the advantage of amplitude embedding which is analyzed to assist in constructing a comparatively balanced tree.
\section{Quantum Fidelity Decision Tree}
In this section, we discuss the quantum state generation from a classical dataset examine the quantum splitting criteria for constructing a decision tree. In order to facilitate the discussion, we first briefly go through the required terminology. In general, a machine-learning dataset, $\mathbb{D}$, consists of a set of features $X$ and corresponding class labels $\mathbb{Y}$ such that $\mathbb{D} = \{X,\mathbb{Y}\}$ where the cardinality of features defines dimensions of the dataset. Further, a dataset contains n-instances represented as $\{(X_1,\mathbb{Y}_1),(X_2,\mathbb{Y}_2), ..., (X_i,\mathbb{Y}_j), ..., (X_n,\mathbb{Y}_n)\}$ where $X_i$ is \textit{i}-th instance of the feature and $\mathbb{Y}_j$ is a corresponding \textit{j}-th instance for the class label. In the following subsections, we will use these notations to describe the quantum splitting criterion and construction of a quantum decision tree.

\subsection{Feature Embedding}
In general, all splitting criteria of a classical decision tree are influenced by the function of a probability distribution of a random event for feature $X_{i}$. The features in a dataset are segregated into two or more classes leading to a probability distribution corresponding to different classes which can be further used to obtain the information gain \cite{quinlan1986induction,quinlan1989inferring}. Although, there are different ways to embed classical data into a quantum state, here, we will emphasize on amplitude embedding as a preferred embedding technique. In general, the amplitude embedding is used with respect to features, however, we use a modified amplitude embedding method to process the classical probability distribution into quantum states as an amplitude value. For this, we construct a quantum state using a combined probability distribution using feature and associated class labels. Algebraically, such a state can be represented as 
\begin{equation}\label{psi}
    \ket{\psi_{i,j}} = \sum_{i,j} p_{j|i}\ket{X_{i} \mathbb{Y}_{j}}
\end{equation}
where $p_{j|i}$ is the probability distribution of class $\mathbb{Y}$ with the occurrence of a random event for feature $X$.
\subsection{Quantum Splitting Criteria}
For a classical decision tree, classical information gain and gini index are used extensively as splitting criteria \cite{quinlan1986induction,breiman2017classification}. In order to represent a composite quantum system and to understand the properties and dynamics of subsystems, density operator can be used as an efficient mathematical tool instead of a wave function representation. 
For example, if the state of a system is specified by $\ket{\psi_i}$ where $i$ is the index for the data point occurring with a probability $p_i$ in an ensemble $\{p_i, \ket{\psi{_i}}\}$, then the spectral decomposition of the corresponding  density operator can be expressed as
\begin{equation}\label{rho}
    \rho = \sum_i p_i\ket{\psi_i}\bra{\psi_i}
\end{equation}
here $\rho$ represents a density operator. Therefore, the density operator of a composite system for feature $X_i$ and corresponding class label $\mathbb{Y}_j$ can be further expressed using Eqs. \ref{psi} as:
\begin{equation}\label{rho_{X_i, y_j}}
    \rho_{X_i\mathbb{Y}_j} = \sum_{i,j} p_{i,j} \ket{\psi_{i,j}}\bra{\psi_{i,j}}
\end{equation}
For the splitting criterion, we use the concept of fidelity using reduced density operators $\rho_{X}$ and $\rho_{\mathbb{Y}}$ for individual subsystems, i.e., feature $X$ and class-label $\mathbb{Y}$, respectively. The proposed splitting criterion for our purpose can therefore be depicted as
\begin{equation}\label{fidelity}
    F(\rho_{X},\rho_{\mathbb{Y}}) = tr(\sqrt{\sqrt{\rho_{X}}\rho_{\mathbb{Y}}\sqrt{\rho_{X}}})
\end{equation}
The fidelity in Eq. \ref{fidelity} is bounded by $0 \leq F(\rho_{X},\rho_{\mathbb{Y}}) \geq 1$, where the zero fidelity corresponds to the maximum distance between the two states or dissimilarity between a feature and the associated class label. Similarly, if the fidelity is unity, it represents the perfect overlap between the two states.   Therefore, we calculate the fidelity for each feature in a dataset $\mathbb{D}$ and then select the maximum fidelity value for splitting the dataset. Algorithm \ref{alg:qf} below depicts the quantum node splitting criteria.  
\begin{algorithm}
    \caption{Quantum node splitting criterion}\label{alg:qf}
    \textbf{Quantum Fidelity} (features X, labels $\mathbb{Y}$);\\
    Choose feature = X and class-label = $\mathbb{Y}$;\\
    dictionary(x) = Unique(x);\\
    \While{i in dictionary(x)}
    {
    compute and store $(p_{j|i})$;
    }
    compute $(\ket{\psi_{i,j}} = \sum_{i,j} p_{j|i}\ket{X_{i} \mathbb{Y}_{j}})$;\\
    Normalize$(\ket{\psi_{i,j}})$;\\
    compute $(\rho_{X\mathbb{Y}}, \rho_{X}, \rho_{\mathbb{Y}})$;\\
    compute $(F = tr(\sqrt{\sqrt{\rho_{X}}\rho_{\mathbb{Y}}\sqrt{\rho_{X}}}))$;\\
    add F in Fidelity list;\\
    Use Grover's search algorithm for j where $ Fidelity_{j}> Fidelity_{max} $;\\
    set $Fidelity_{max} = Fidelity_{j}$;\\
    Return $R_I = X^{k} Fidelity_{max}$;\\
\end{algorithm}
As per algorithm \ref{alg:qf}, we first compute probability distribution for a class-label conditioning on each random event in a feature. Using this probability distribution for each event, we further construct a quantum state to compute  $\rho_{X\mathbb{Y}}$, $\rho_{X}$, and $\rho_{\mathbb{Y}}$. For the splitting criterion, our algorithm used reduced density operators $\rho_{X}$, and $\rho_{\mathbb{Y}}$ to evaluate fidelity for each feature.  Finally, we utilize Grover's search algorithm to extract the maximum fidelity value among all evaluated values. The feature corresponding to the maximum fidelity will be used for splitting the dataset.

\subsection{Constructing Quantum-Classical Decision Tree}
We now proceed to construct the full decision tree, which is based on quantum splitting criteria. For a dataset $\mathbb{D}$ containing $X_{i}^{k}$ features and $\mathbb{Y}_{j}$ labels, we embed amplitude values into quantum Hilbert space using the feature mapping. Further, using Algorithm \ref{alg:qdt}, we create a node by analysing the following three conditions: 
\begin{enumerate}
    \item if attributes in a dataset are empty then label the root node ($R$) with majority class-label in dataset
    \item if all instances belong to the same class-label then assign that class-label to node
    \item if instances are empty then label of the node is decided by the label of previously seen examples
\end{enumerate}
Else if all the above conditions are not satisfied then we compute the fidelity for each feature using Algorithm \ref{alg:qf} and choose an attribute with the highest value.

\begin{algorithm}
\caption{Quantum-Classical Decision Tree}\label{alg:qdt}
\textbf{Quantum-Classical Decision Tree($\mathbb{D}$, X, $\mathbb{Y}$)};\\
Create Empty node $R$;\\
\eIf{$X_{i} == 0$, $\forall$ $i$ $\in$ $n$}
{
Return $\mathbb{Y}$ of $max$$(X_{i-1})$, for $l$;
}{
    \If {$\forall i \in n$, $X_{i}$ $\in$ $\mathbb{Y}$}
    {
        Return $\mathbb{Y}$, for $l$;
    }

    \If {$\forall$ $k$ $\in$ $d$, $X^{k}$ empty}
    {
    $l$ = $\mathbb{Y}_{j}$ of $max(X_{i})$;
    }
}

$R_I$ = \textbf{Quantum Fidelity(X, $\mathbb{Y}$)};\\
Return tree ($R$ to $R_I$);
\end{algorithm}

In the above algorithm, $l$ is the leaf node to which we assign a class-label $\mathbb{Y}$ as per the mentioned conditions and $R_I$ is the descendent node of $R$.
\section{Numerical Analysis}
In order to demonstrate the effectiveness of the proposed fidelity-based quantum decision tree algorithm, we first analyse our  numerical results for a dataset shown in the table below:\par
\begin{figure}
    \centering
    \includegraphics[width = 0.48\textwidth]{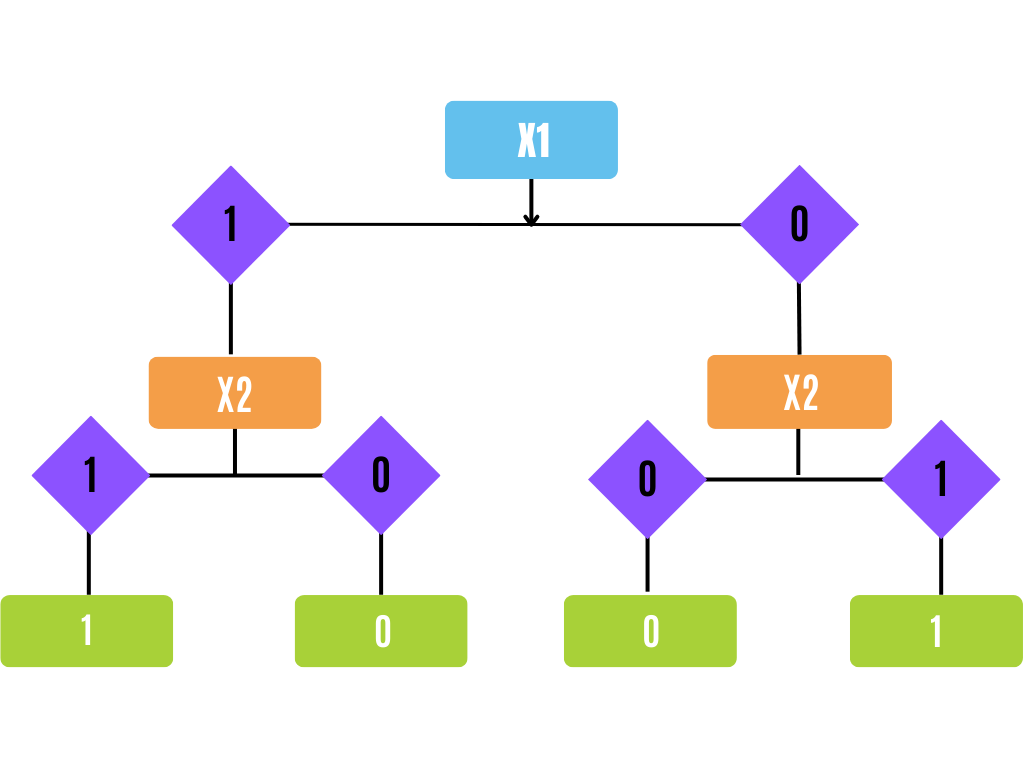}
    \caption{A fully developed balanced quantum decision tree}
    \label{fig:qft}
\end{figure}
\begin{figure}
    \centering
    \includegraphics[width = 0.48\textwidth]{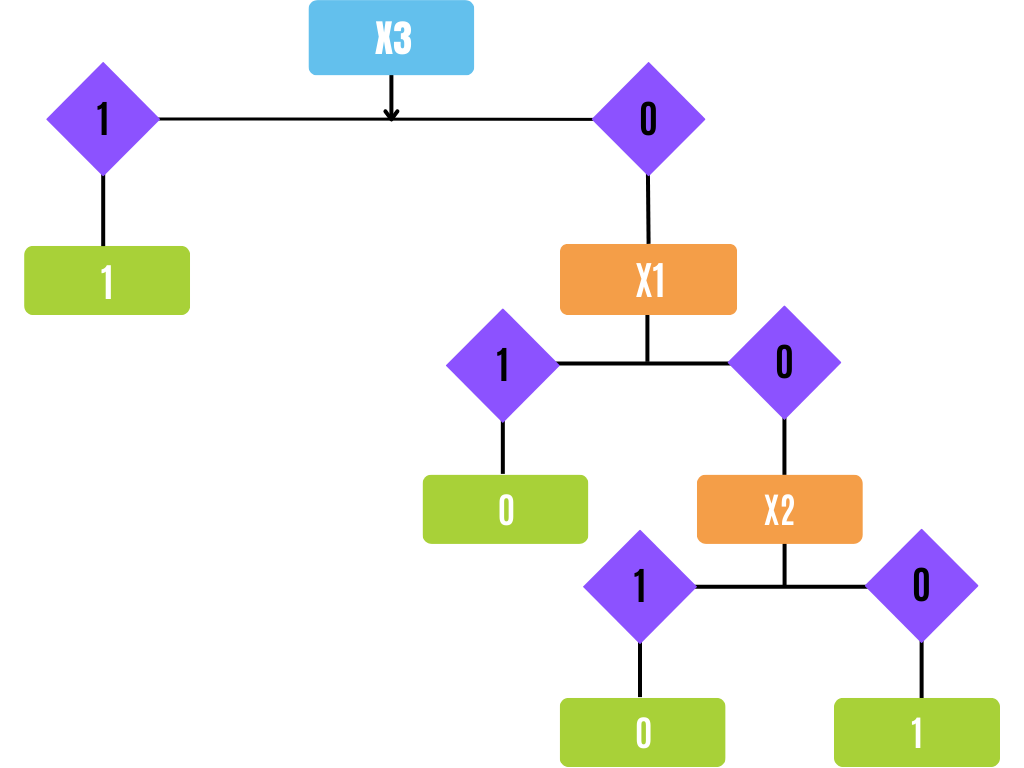}
    \caption{A fully developed classical decision tree}
    \label{fig:cig}
\end{figure}

\begin{table}[h]
\centering
    \begin{tabular}{ c | c | c | c }
        X1 &  X2 & X3 & Y\\
        \hline
        1 & 1&1&1\\
        0&1&0&0\\
        1&0&1&0\\
        0&0&1&1\\
    \end{tabular}
     \caption{The binary valued dataset with $X1$, $X2$ and $X3$ features and $Y$ label}
     \label{binary dataset}
\end{table}
For the dataset in Table \ref{binary dataset}, we first evaluate three conditions as specified in Algorithm \ref{alg:qdt}. Since none of the three conditions satisfy, we proceed towards selecting an attribute for splitting. Table \ref{binary dataset} clearly demonstrates that the set of random events for all features are either ${0}$ or ${1}$; and similarly class-labels are also either ${0}$ or ${1}$. Therefore, using Table \ref{binary dataset}, we express a two-qubit quantum state for each feature and corresponding class label as 
\begin{equation}\label{eqnor.psi_x1,x2,x3}
\begin{aligned}
    \ket{\psi_{X1}} = \frac{1}{2} \ket{00} + \frac{1}{2} \ket{01} + \frac{1}{2} \ket{10} + \frac{1}{2} \ket{11}\\
    \ket{\psi_{X2}} = \frac{1}{2} \ket{00} + \frac{1}{2} \ket{01} + \frac{1}{2} \ket{10} + \frac{1}{2} \ket{11}\\
    \ket{\psi_{X3}} = \frac{1}{\sqrt{6}}\{\ket{00} + \ket{10} + 2 \ket{11}\}
\end{aligned}
\end{equation}
Eq. \ref{eqnor.psi_x1,x2,x3} shows that the occurrence of the events in states $\ket{\psi_{X1}}$ and $\ket{\psi_{X2}}$ are equally probable. Alternately, one can visualize that a splitting with respect to the feature $X1$ or $X2$ will result in a balanced decision tree. For constructing the decision tree, we now compute the two qubit density operators using Eq. \ref{eqnor.psi_x1,x2,x3} and then further evaluate fidelity with  $\rho_{X}$, and $\rho_{\mathbb{Y}}$, the reduced density operators as described in Algorithm \ref{alg:qf},
\begin{table}[]
    \centering
    \begin{tabular}{c|c|c}
        Features & Fidelity & CIG \\
        \hline
        $X1$ & 1 & 0\\
        $X2$ & 1 & 0\\
        $X3$ & 0.98 & 0.31
    \end{tabular}
    \caption{The \textbf{Quantum Fidelity} and Classical Information Gain Values corresponding to $X1$, $X2$ and $X3$ features}
    \label{tab:fidelity_values}
\end{table}
\begin{figure}
    \centering
    \includegraphics[width=0.48\textwidth]{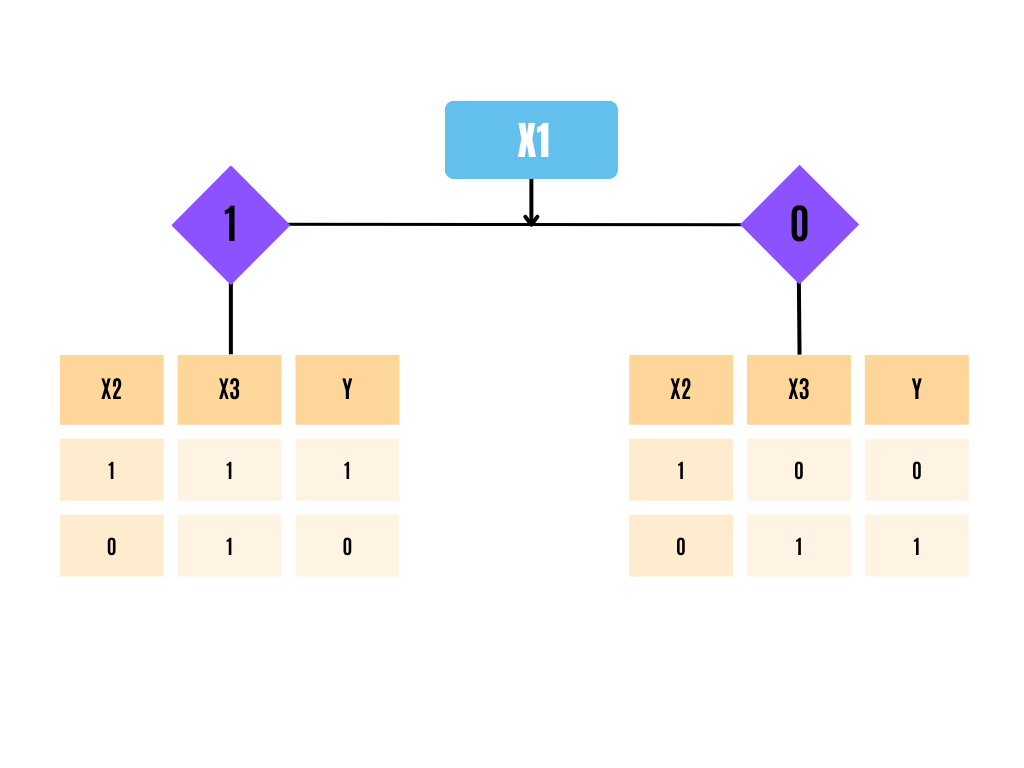}
    \caption{Decision tree developed using
Quantum Fidelity Decision Tree Algorithm, splitting over the X1
feature}
    \label{fig:quantum_step1}
\end{figure}
\begin{figure}
    \centering
    \includegraphics[width=0.48\textwidth]{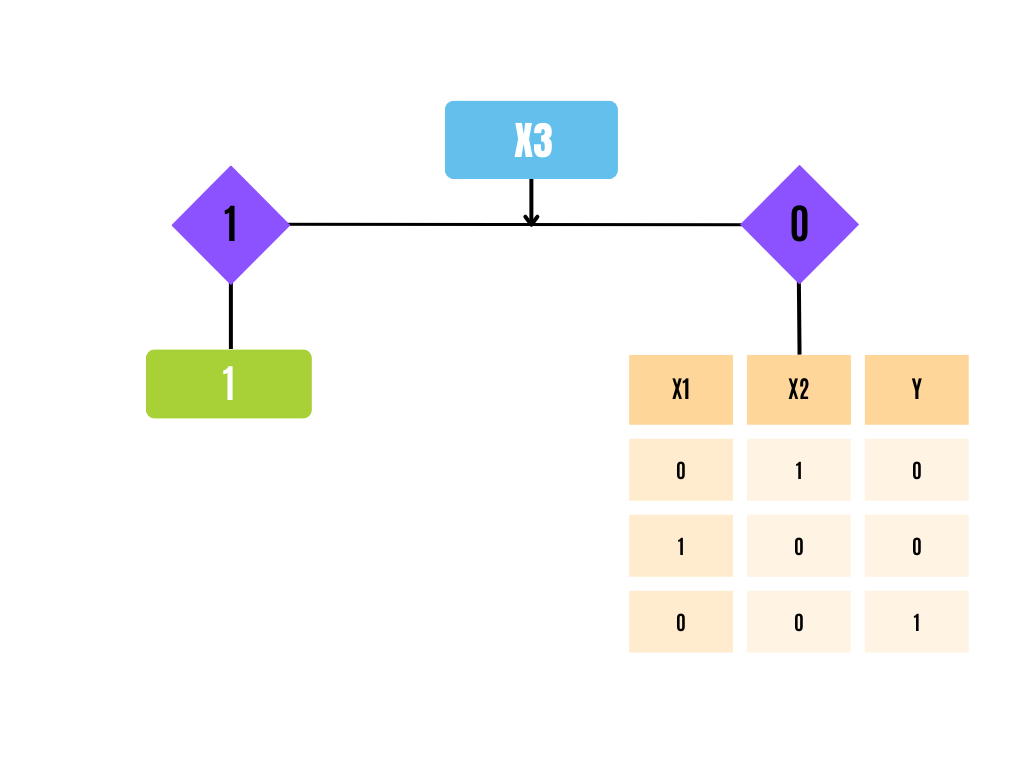}
    \caption{Decision tree developed using
Classical Information Gain Algorithm, splitting over the X3
feature}
    \label{fig:classical_step1}
\end{figure}
Table: \ref{tab:fidelity_values} demonstrates fidelity and classical information gain corresponding to each feature. 
For the splitting criterion, fidelities corresponding to features $X1$ and $X2$ are the maximum therefore either of the features can be used as a splitting node- a quantum decision tree. Whereas, for the classical information gain, the highest value is achieved by $X3$ and therefore the feature $X3$ can be used as a splitting node- a classical decision tree. Figs. \ref{fig:quantum_step1} and \ref{fig:classical_step1} represent the constructed quantum and classical decision trees for the selected root nodes, respectively. The process recursively repeats itself until it meets any of the three conditions specified in Algorithm \ref{alg:qdt}. Hence, Figs. \ref{fig:qft} and \ref{fig:cig} demonstrate full-grown quantum and classical decision trees, respectively. As discussed earlier, the quantum decision tree being splitted at $X1$ or $X2$ is a completely balanced tree and the classical decision tree being splitted at $X3$ due to its greedy nature is an imbalanced tree. For analyzing the importance and effectiveness of the proposed algorithm, we further evaluate our criterion using different publicly available datasets in the next section. 

\section{Simulation Results}
In this section, we proceed to analyze the effectiveness of fidelity as a quantum splitting criterion. For this, we use datasets as indicated in Table \ref{tab:datasets} \cite{kellyuci}.
\begin{table}[]
    \centering
    \begin{tabular}{c|c|c}
    \hline
        Name of dataset & \# of instances & dimesions\\
        \hline
         Haberman's cancer survival dataset & 306 & 3\\
         Wheat seed dataset & 140 & 7\\
         Wisconsin breast cancer dataset & 699 & 9
    \end{tabular}
    \caption{Description of the datasets}
    \label{tab:datasets}
\end{table}
Table \ref{tab:datasets} describes datasets in terms of dimensions (number of features) and size of that particular dataset where Haberman's cancer survival dataset \cite{misc_haberman's_survival_43} analyzes whether a patient will survive for 5 years or will die within 5 years based on the age factor, number of nodes and year of operation performed. Similarly, Wisconsin breast cancer dataset \cite{misc_breast_cancer_wisconsin_(original)_15} preditcs the breast tissues as benign or malignant based on nine different features , i.e., clump thickness, uniformity of cell size, uniformity of cell shape, marginal adhesion, single epithelial cell size, bare nuclei, bland chromatin, normal nucleoli, and mitoses. The Wheat seed dataset \cite{misc_seeds_236} further classifies Kama and Canadian seeds based on area, perimeter, compactness, length of kernel, width of kernel, asymmetry coefficient, and length of the kernel groove. \par
Out of the three datasets, Wisconsin breast cancer and Haberman's cancer survival datasets are highly imbalanced for two classes with a ratio of $65.52$ vs $34.48$ and $73.53$ vs $26.47$, respectively. On the other hand, Wheat seed dataset is completely balanced. 
 \par
\begin{figure}
    \centering
    \includegraphics[width =0.48 \textwidth]{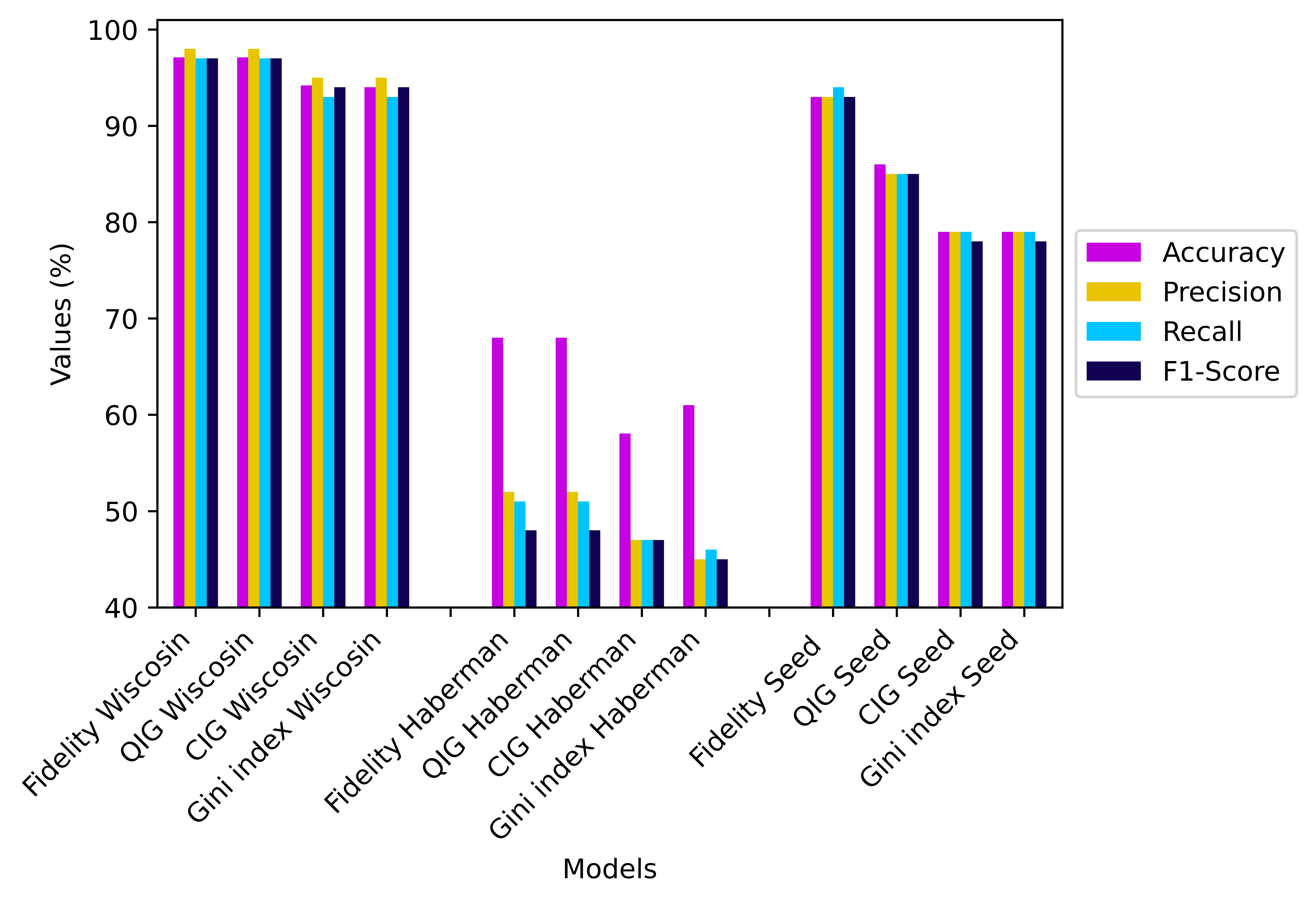}
    \caption{Performance measure and comparison of proposed quantum decision trees and classical decision tree algorithms, here QIG refers to quantum information gain and CIG is to classical information gain}
    \label{fig:em_models}
\end{figure}
\begin{equation}\label{accuracy}
    Accuracy = \frac{TP+TN}{TP+FP+FN+TN} 
\end{equation}
\begin{equation}\label{precision}
\begin{aligned}
    & Precision(macroavg) = \\
    & \frac{TP_{class0} + TP_{class1}}{TP_{class0}+TP_{class1}+FP_{class0} + FP_{class1}}
\end{aligned}
\end{equation}
\begin{equation}\label{recall}
\begin{aligned}
        & Recall(macroavg) = \\
        & \frac{TP_{class0} + TP_{class1}}{TP_{class0}+TP_{class1}+FN_{class0} + FN_{class1}}
\end{aligned}
\end{equation}
\begin{equation}\label{f1-score}
    F1-Score = 2*\left(\frac{Precision * Recall}{Precision + Recall}\right)
\end{equation}
In order to evaluate the performance of the proposed quantum splitting criteria (fidelity and quantum information gain) as against the classical splitting criteria (classical information gain and gini index), we train the data on $90 \%$ of overall dataset and test the data on $10 \%$ of overall dataset. The algorithms are assessed in terms of four important metrics, namely accuracy, precision, recall, and f1-score as represenetd in Eqs. \ref{accuracy}, \ref{precision}, \ref{recall}, and \ref{f1-score}, respectively where TP stands for True Positive, TN stands for True Negative, FP stands for False Positive and FN stands for False Negative. Since Wisconsin breast cancer and Haberman's cancer survival datasets are imbalanced,  precision, recall and f1-score metrics are preferred instead of accuracy in dealing with these two datasets \cite{juba2019precision}.\par
The efficiency of the proposed fidelity based quantum splitting criterion is evaluated by comparing it to classical information gain, gini index, and quantum information gain splitting criteria. Fig. \ref{fig:em_models} demonstrates the advantages of using proposed quantum splitting criteria based on fidelity and quantum information gain (QIG). Betweeen gini index and CIG, gini index is found to be more efficient for Haberman's cancer survival dataset in terms of accuracy; however, for precision and recall metrics, the classical information gain performs better than gini index. Whereas, for Wisconsin and wheat seed datasets, the efficiency of classical splitting criteria is same. 
For quantum splitting criteria, the efficiency of fidelity and quantum information gain based splitting criteria is exactly the same for Wisconsin breast cancer and Haberman's cancer survival datasets. Interestingly, the performance of quantum splitting criteria is significantly better than classical splitting criteria  in terms of all evaluation metrics for all datasets. For, the Wheat seed dataset, our results found fidelity based splitting criterion to be significantly better than the quantum information gain splitting criterion for all metrics. \par
\begin{table}[]
    \centering
    \begin{tabular}{m{5em}|m{3cm}|m{1.5cm}|m{1cm}|m{1cm}}
    \hline
        Methods & Datasets & Specificity & PPV & NPV \\
        \hline
        \multirow{2}{6em}{Fidelity}  & Haberman's Cancer Dataset & 90.90\% & 33.33\% & 71.42\%\\
        \cline{2-5}
        & Wisconsin's Breast Cancer Dataset & 100\% & 100\% & 95.12\%\\
        \hline
        \multirow{2}{6em}{Quantum Information Gain} & Haberman's Cancer Dataset & 90.90\% & 33.33\% & 71.42\%\\
        \cline{2-5}
         & Wisconsin's Breast Cancer Dataset & 100\% & 100\% & 95.12\%\\
         \hline
        \multirow{2}{6em}{Classical Information Gain} & Haberman's Cancer Dataset & 72.72\% & 25\% & 69.56\%\\
        \cline{2-5}
         & Wisconsin's Breast Cancer Dataset & 100\% & 100\% & 90.69\%\\ 
         \hline
        \multirow{2}{6em}{Gini Index } & Haberman's Cancer Dataset & 81.81\% & 20\% & 69.23\%\\
        \cline{2-5}
         & Wisconsin's Breast Cancer Dataset & 100\% & 100\% & 90.69\%\\
    \end{tabular}
    \caption{Detailed evaluation of imbalance dataset and comparison of all splitting criteria of quantum and classical decision tree algorithms}
    \label{tab:medical_dia}
\end{table}
\begin{equation}\label{specificity}
    Specificity = \frac{TN}{TN+FP}
\end{equation}
\begin{equation}\label{ppv}
    Positive Predictive Value (PPV) = \frac{TP}{TP+FP}
\end{equation}
\begin{equation}\label{npv}
    Negative Predictive Value (NPV) = \frac{TN}{TN+FN}
\end{equation}
Considering the importance of true positive and true negative results in medical datasets, these are also analyzed based on some additional metrics such as specificity, positive predictive value and negative predictive value as represented in Eqs. \ref{specificity}, \ref{ppv}, and \ref{npv}, respectively. Here, specificity represents true negative result out of all negative instance;, PPV represents true positive result out of actual positive and predicted positive; and NPV represents for true negative outcomes out of true negative and predictive negative. Table \ref{tab:medical_dia} shows effects of different splitting criteria on imbalanced datasets. Clearly, for the Wisconsin breast cancer dataset, quantum splitting criteria show highest true negative predictions compared to classical splitting criteria suggesting that the model can classify the unknown data with high degree of accuracy. Surprisingly, for the Haberman's cancer survival dataset, the fidelity and quantum information gain splitting criteria significantly exceed classical splitting criteria in specificity, PPV and NPV, which shows that our models can efficiently and accurately classify the data into the two classes. For classical splitting criteria,  gini leads to a better result in specificity in comparison to classical information gain. 
\section{Conclusion}
In this work, we have proposed and analyzed a quantum splitting criterion, i.e., fidelity to construct a decision tree. For the proposed criterion, we have efficiently utilized the probability distribution from classical dataset to generate a quantum state, which is further used to compute fidelity for each feature. The numerical analysis showed that the fidelity splitting criterion selects the feature with a uniform probability distribution. This further assists in obtaining a balanced and more accurate decision tree. Our results demonstrated that the proposed fidelity-based criterion is able to provide a significant difference in terms of all evaluation metrics even for an imbalanced dataset. Furthermore, we also used quantum information gain as a criterion to achieve significantly better results in comparison to classical algorithms for all datasets. For comprehensive analysis, we have examined the efficiency of all quantum and classical splitting methods on precision and recall values, which play a crucial role in medical datasets. The obtained results clearly demonstrate the advantages of quantum splitting criteria over classical splitting criteria.

\bibliography{main}
\bibliographystyle{unsrt}
\end{document}